\documentclass[12pt]{iopart}

\usepackage{graphicx}
\usepackage{longtable}
\usepackage{tabularx}
\begin{document}

\title[Insight into high-entropy effect in body-centered cubic superconducting alloys]{Insight into high-entropy effect in body-centered cubic superconducting alloys}

\author{Hanabusa Senga$^1$, Yuto Watanabe$^2$, Fubuki Iwase$^1$, Ryo Masuda$^1$, Daichi Kawahara$^1$, Toshiki Haruyama$^1$, Terukazu Nishizaki$^3$, Yoshikazu Mizuguchi$^2$, and Jiro Kitagawa$^1$}

\address{$^1$ Department of Electrical Engineering, Faculty of Engineering, Fukuoka Institute of Technology, 3-30-1 Wajiro-higashi, Higashi-ku, Fukuoka, 811-0295, Japan}
\address{$^2$ Department of Physics, Tokyo Metropolitan University, Hachioji, 192-0397, Japan}
\address{$^3$ Department of Electrical Engineering, Faculty of Science and Engineering, Kyushu Sangyo University, 2-3-1 Matsukadai, Higashi-ku, Fukuoka, 813-8503, Japan}
\ead{j-kitagawa@fit.ac.jp}

\begin{abstract}
We have characterized the superconducting critical temperature ($T_\mathrm{c}$), the Debye temperature ($\theta_\mathrm{D}$), the electronic specific heat coefficient, and the Vickers microhardness of HfNbTiVZr, NbTiZr, HfNbTi, HfNbZr, and HfNbTa, all possessing a body-centered cubic (bcc) structure.
By compiling a comparable dataset for other equiatomic quinary bcc high-entropy alloy (HEA) superconductors, we have examined the validity of the hypothesis regarding the high-entropy effect in bcc HEA superconductors, as proposed in our previous work (J. Alloys Compd. 924 (2022) 166473). 
This hypothesis attributes the observed negative correlation between the electron-phonon coupling constant ($\lambda_\mathrm{e-p}$) and $\theta_\mathrm{D}$ to a reduced phonon lifetime at higher $\theta_\mathrm{D}$, arising from the uncertainty principle in highly disordered quinary alloys. 
However, a pronounced change in this negative correlation is not evident in equiatomic ternary alloys with a lower degree of atomic disorder, thereby providing limited support for the hypothesis.
Alternatively, by assembling the full dataset of bcc alloys spanning binary through senary systems, we have identified a universal negative correlation between $\lambda_\mathrm{e-p}$ and $\theta_{D}$. 
This result would be useful for the materials design of bcc superconducting alloys. 
We further propose that the Vickers microhardness offers an alternative means to evaluate $\theta_{D}$ and may serve as a rapid screening metric for identifying bcc alloys with desired properties.
\end{abstract}

\noindent{\it keywords}: High-entropy alloys, Superconductivity, Debye temperature, body-centered cubic

%
%
%
%
%

\section{Introduction}
Traditional alloys typically consist of a base element added by small amounts of additional elements. 
However, this paradigm has shifted with the advent of high-entropy alloys (HEAs)\cite{Mishra:MSEA2021}. 
The design principle of HEAs relies on configurational entropy, whose increase contributes to the thermodynamic stability of the alloy. 
Consequently, HEAs generally comprise more than four distinct elements in nearly equiatomic proportions\cite{Yan:MMTA2021}. 
HEAs have demonstrated superior functionalities, including outstanding mechanical performance, irradiation resistance, energy storage capability, thermoelectric properties, and soft ferromagnetism, among others\cite{Li:PMS2021,Moschetti:JNM2022,Marques:EES2021,Jiang:Science2021,Chaudhary:MT2021}. 
Notably, their multifunctionality—generally unattainable in conventional alloys—has recently attracted considerable interest\cite{Han:NRM2024}. 
For example, the simultaneous realization of ductility and high strength, or the coexistence of soft ferromagnetism and corrosion resistance, has been reported\cite{Shi:NC2019,Duan:SCM2023}.

Superconductivity also represents a significant functionality in electronic HEA materials\cite{Kitagawa:book}. 
Recent studies on HEA superconductors have revealed that their critical current densities are comparable to those of commercial Nb–Ti alloys\cite{Jung:NC2022,Gao:APL2022,Kim:JMST2024,Seki:JSNM2024,Kitagawa:EPJB2025}. 
Since many HEAs exhibit remarkable irradiation resistance\cite{Moschetti:JNM2022}, HEA superconductors hold promise for multifunctional superconducting wires applicable to nuclear fusion reactors and aerospace technologies. 
From a fundamental perspective, extensive research has been devoted to elucidating the high-entropy effect on superconducting properties. 
One notable phenomenon is the cocktail effect, wherein physical properties are enhanced beyond the composition-weighted averages predicted by the mixture rule. 
In body-centered cubic (bcc) and hexagonal close-packed HEAs, the cocktail effect has been investigated in relation to the superconducting critical temperature ($T_\mathrm{c}$)\cite{Kozelj:PRL2014,Ishizu:RINP2019,Liu:PRM2024}, where the experimental $T_\mathrm{c}$ markedly exceeds the predicted value. 
Another example is the violation of Matthias rule\cite{Kitagawa:book,Kitagawa:Metals2020,Idczak:PRB2025}, which highlights the strong correlation between the valence electron concentration per atom (VEC) and $T_\mathrm{c}$ in binary or ternary transition-metal superconductors. 
In traditional bcc alloys with VEC values between 3.5 and 5.5, $T_\mathrm{c}$ exhibits a pronounced peak near a VEC of 4.6. 
However, many bcc HEAs deviate from Matthias rule, an anomaly attributed to the elemental makeup of these alloys\cite{Kitagawa:Metals2020,Idczak:PRB2025,Rohr:PRM2018}.

Our group has previously proposed a high-entropy effect in bcc HEA superconductors\cite{Kitagawa:JALCOM2022}. 
In that work, we investigated the superconducting properties of HfMoNbTiZr, in which the Ta element in the widely studied bcc HEA Hf–Nb–Ta–Ti–Zr is replaced with the bcc element Mo so as to preserve the bcc structure.
A comparative analysis of superconducting properties among four equiatomic quinary bcc alloys\cite{Vrtnik:JALCOM2017,Marik:JALCOM2018,Sarkar:IM2022,Kitagawa:JALCOM2022}—HfNbTaTiZr, HfNbReTiZr, HfNbTaTiV, and HfMoNbTiZr—provides a unique perspective, as these alloys possess identical maximum configurational entropy, thereby enabling a direct examination of the high-entropy effect. 
This study suggested that $T_\mathrm{c}$ decreases with a reduction in the electron–phonon coupling constant ($\lambda_\mathrm{e-p}$), which is negatively correlated with the Debye temperature ($\theta_{D}$). 
We posited that this negative correlation arises from shortened phonon lifetimes at higher $\theta_{D}$, a manifestation of the uncertainty principle relating phonon energy uncertainty to lifetime. 
In highly disordered atomic systems, such as equiatomic quinary HEAs, phonon broadening is pronounced\cite{Kormann:npjCM2017} and becomes increasingly significant with larger $\theta_{D}$. 
Thus, higher $\theta_{D}$ leads to shorter phonon lifetimes through the uncertainty principle, thereby weakening $\lambda_\mathrm{e-p}$. 
If this hypothesis holds, the negative correlation between $\lambda_\mathrm{e-p}$ and $\theta_{D}$ should be substantially altered in equiatomic ternary bcc alloys, where atomic disorder is greatly reduced. 
To validate this hypothesis, a comparative study of equiatomic quinary and ternary bcc alloys is essential, with the collection of as extensive a dataset as possible. 
Importantly, several additional equiatomic quinary superconducting bcc HEAs have been reported since our earlier work\cite{Jangid:APL2024,Dong:SM2023,Sharma:PRM2025,Zeng:AQT2023,Jangid:PRM2025,Nowak:AM2025}, increasing the number of known superconductors to ten.

In this study, our primary objective was to compile a comprehensive dataset of $T_\mathrm{c}$ and $\theta_\mathrm{D}$ for the ten equiatomic quinary bcc HEAs. 
Among these, HfNbTiVZr lacked a reported value for $\theta_\mathrm{D}$, which we have determined in this work. 
For many equiatomic ternary bcc alloys, reliable values of $\theta_\mathrm{D}$ remain scarce.
Accordingly, we have carried out systematic investigations on equiatomic ternary alloys by removing two elements from the representative equiatomic HfNbTaTiZr and consistently obtained $T_\mathrm{c}$ and $\theta_\mathrm{D}$ for NbTiZr, HfNbTi, HfNbZr, and HfNbTa, all of which exhibit single-phase bcc structures. 
With these datasets complete, we examine the negative correlation between $\lambda_\mathrm{e-p}$ and $\theta_\mathrm{D}$ across quinary and ternary bcc alloys. 
After confirming that this examination provides limited support for our hypothesis, we evaluated the correlation between $\lambda_\mathrm{e-p}$ and $\theta_\mathrm{D}$ by compiling the full dataset reported for bcc alloys spanning binary to senary systems. 
We identified a universal negative correlation between $\lambda_\mathrm{e-p}$ and $\theta_\mathrm{D}$, irrespective of the degree of atomic disorder. 
Finally, we propose the potential of Vickers microhardness as a rapid screening method for identifying bcc superconducting alloys with desired properties.

\section{Materials and Methods}
Polycrystalline specimens of HfNbTiVZr, NbTiZr, HfNbTi, HfNbZr, and HfNbTa were synthesized using a homemade arc furnace under an Ar atmosphere. 
The starting atomic ratios were Hf:Nb:Ti:V:Zr=1:1:1:1:1 for HfNbTiVZr, Nb:Ti:Zr=1:1:1 for NbTiZr, Hf:Nb:Ti=1:1:1 for HfNbTi, Hf:Nb:Zr=1:1:1 for HfNbZr, and Hf:Nb:Ta=1:1:1 for HfNbTa, respectively.
Each ingot was inverted and remelted multiple times, followed by quenching on a water-cooled Cu hearth.

Room-temperature X-ray diffraction (XRD) patterns were obtained with an X-ray diffractometer (XRD-7000L, Shimadzu) employing Cu-K$\alpha$ radiation in the Bragg–Brentano geometry. 
Thin slabs sectioned from the ingots were used owing to the difficulty of preparing fine powders. 
Scanning electron microscopy (SEM) images of polished surfaces were acquired using field-emission scanning electron microscopy (FE-SEM; JSM-7100F, JEOL). 
The local chemical composition of each region was determined with an energy-dispersive X-ray (EDX) spectrometer attached to the FE-SEM, and elemental mapping was likewise obtained with the EDX system.

The temperature dependence of the dc magnetization, $M$($T$), was measured using a SQUID magnetometer (MPMS3, Quantum Design). 
Specific heat, $C_\mathrm{p}$, was determined by the thermal-relaxation technique using a Quantum Design PPMS apparatus. 
Electrical resistivity, $\rho$, in the temperature range of 3 K to room temperature, was measured by a dc four-probe method with a custom-built setup housed in a GM refrigerator (UW404, Ulvac Cryogenics). 
Vickers microhardness was evaluated under a load of 300 g applied for 10 s using a Shimadzu HMV-2T microhardness tester.

\begin{figure}
\begin{center}
\includegraphics[width=0.55\linewidth]{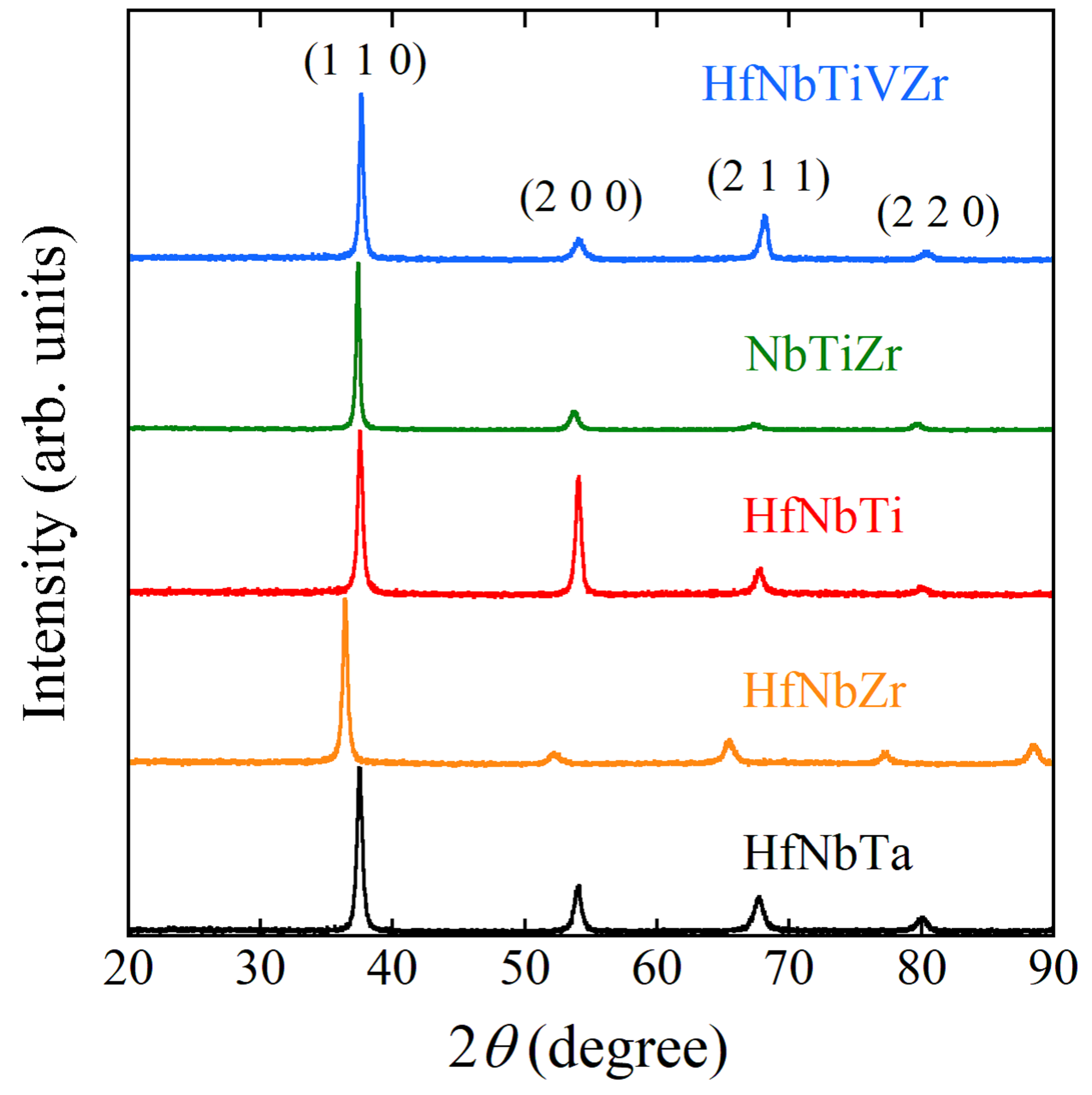}
\caption{\label{fig1}XRD patterns of HfNbTiVZr, NbTiZr, HfNbTi, HfNbZr, and HfNbTa. The origin of each XRD pattern is vertically offset for the clarity.}
\end{center}
\end{figure}

\begin{figure}
\begin{center}
\includegraphics[width=0.9\linewidth]{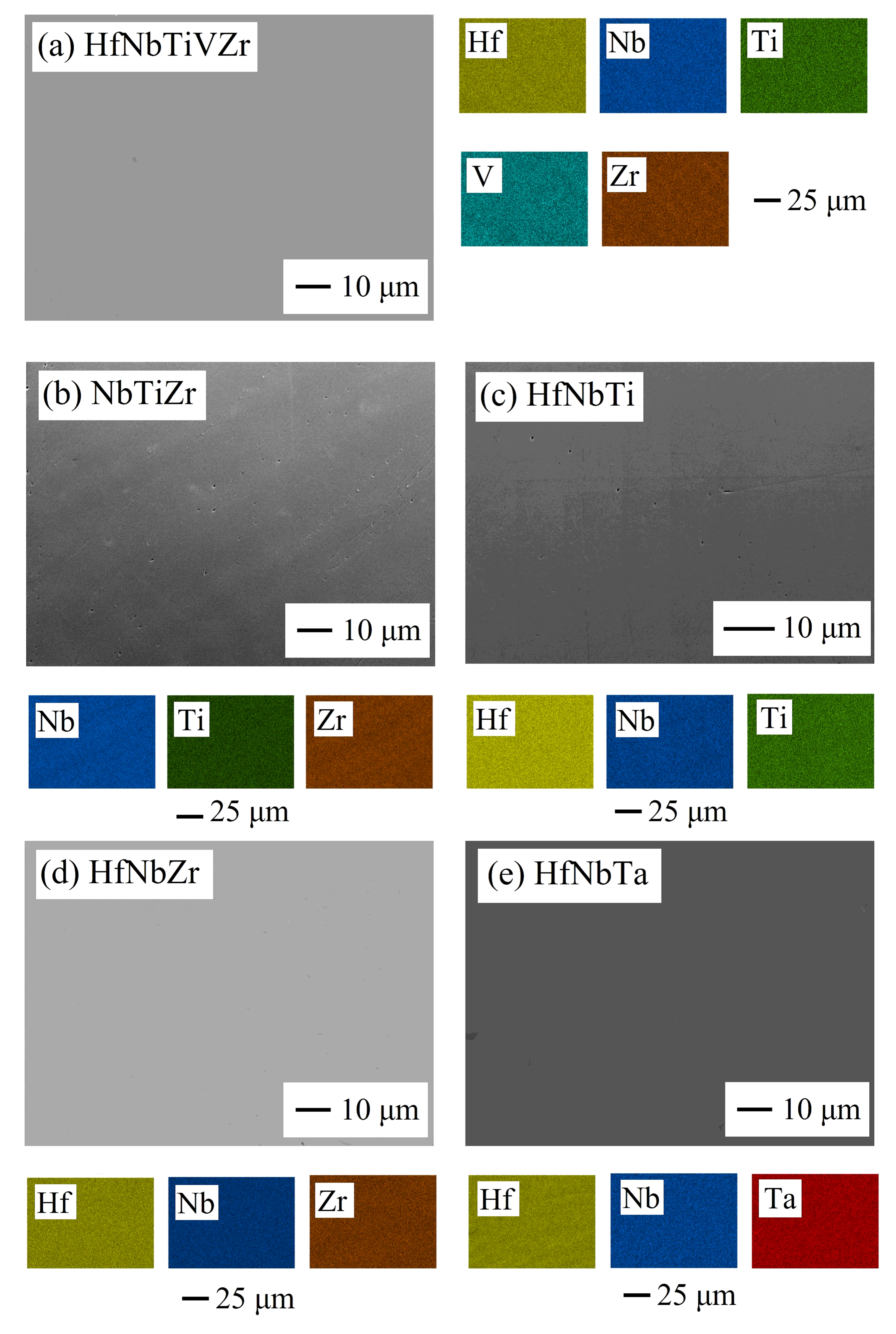}
\caption{\label{fig2}SEM images and corresponding elemental mappings of (a) HfNbTiVZr, (b) NbTiZr, (c) HfNbTi, (d) HfNbZr, and (e) HfNbTa.}
\end{center}
\end{figure}

\section{Results and Discussion}
Figure \ref{fig1} presents the XRD patterns of HfNbTiVZr, NbTiZr, HfNbTi, HfNbZr, and HfNbTa.
All diffraction peaks can be assigned to the Miller indices of the bcc structure, as indicated in the figure. 
The refined lattice parameters are 3.377(3) \AA, 3.400(2) \AA , 3.387(1) \AA, 3.490(3) \AA, and 3.388(1) \AA\hspace{1 mm} for HfNbTiVZr, NbTiZr, HfNbTi, HfNbZr, and HfNbTa, respectively. 
No secondary phases are detected in the SEM images shown in Figs.\hspace{1 mm}\ref{fig2}(a)-(e), which also display homogeneous elemental distributions. 
The chemical compositions determined by EDX are Hf$_{21.4(5)}$Nb$_{20.2(6)}$Ti$_{19.5(3)}$V$_{19.7(9)}$Zr$_{19.1(1)}$ for HfNbTiVZr, Nb$_{34.0(7)}$Ti$_{33.3(4)}$Zr$_{32.7(6)}$ for NbTiZr, Hf$_{34.6(3)}$Nb$_{33.8(5)}$Ti$_{31.6(2)}$ for HfNbTi, Hf$_{33.9(5)}$Nb$_{33.9(9)}$Zr$_{32.2(5)}$ for HfNbZr, and Hf$_{35.3(9)}$Nb$_{33.4(4)}$Ta$_{31.3(6)}$ for HfNbTa, demonstrating satisfactory agreement with the nominal compositions.

Here, we discuss the reliability of the five alloys investigated. 
The ideal lattice parameter is estimated using the hard-sphere model and the chemical compositions obtained from EDX analysis. 
The hard-sphere model of bcc structure yields a lattice parameter expressed as 2.309$\bar{r}$, where $\bar{r}$ represents the composition-weighted atomic radius. 
By employing atomic radii of 1.5775 \AA \hspace{1mm} for Hf, 1.429 \AA \hspace{1mm} for Nb, 1.43 \AA \hspace{1mm} for Ta, 1.4615 \AA \hspace{1mm} for Ti, 1.316 \AA \hspace{1mm} for V, and 1.6025 \AA \hspace{1mm} for Zr\cite{Miracle:AM2017}, the calculated lattice parameters are 3.413 \AA, 3.456 \AA , 3.442 \AA, 3.545 \AA, and 3.421 \AA\hspace{1 mm} for HfNbTiVZr, NbTiZr, HfNbTi, HfNbZr, and HfNbTa, respectively. 
The difference between the experimental lattice parameter and that predicted by the hard-sphere model is denoted as the lattice deviation and defined as $(a_\mathrm{exp}-a_\mathrm{hard})/a_\mathrm{hard}\times 100$, where $a_\mathrm{exp}$ and $a_\mathrm{hard}$ correspond to the experimental and calculated lattice parameters, respectively\cite{Kitagawa:MTC2024}. 
The deviation values are -1.1 \%, -1.6 \%, -1.6 \%,  -1.6 \%, and -1.0 \% for HfNbTiVZr, NbTiZr, HfNbTi, HfNbZr, and HfNbTa, respectively. 
These values are comparable to those of other single-phase bcc HEA superconductors, such as HfMoNbTiZr\cite{Kitagawa:JALCOM2022} (lattice deviation: -1.7 \%) and (Ti$_{35}$Hf$_{25}$)(Nb$_{25}$Ta$_{5}$)Re$_{10}$\cite{Hattori:JAMS2023} (lattice deviation: -2.3 \%), thereby supporting the single-phase nature of HfNbTiVZr, NbTiZr, HfNbTi, HfNbZr, and HfNbTa.

\begin{figure}
\begin{center}
\includegraphics[width=1.0\linewidth]{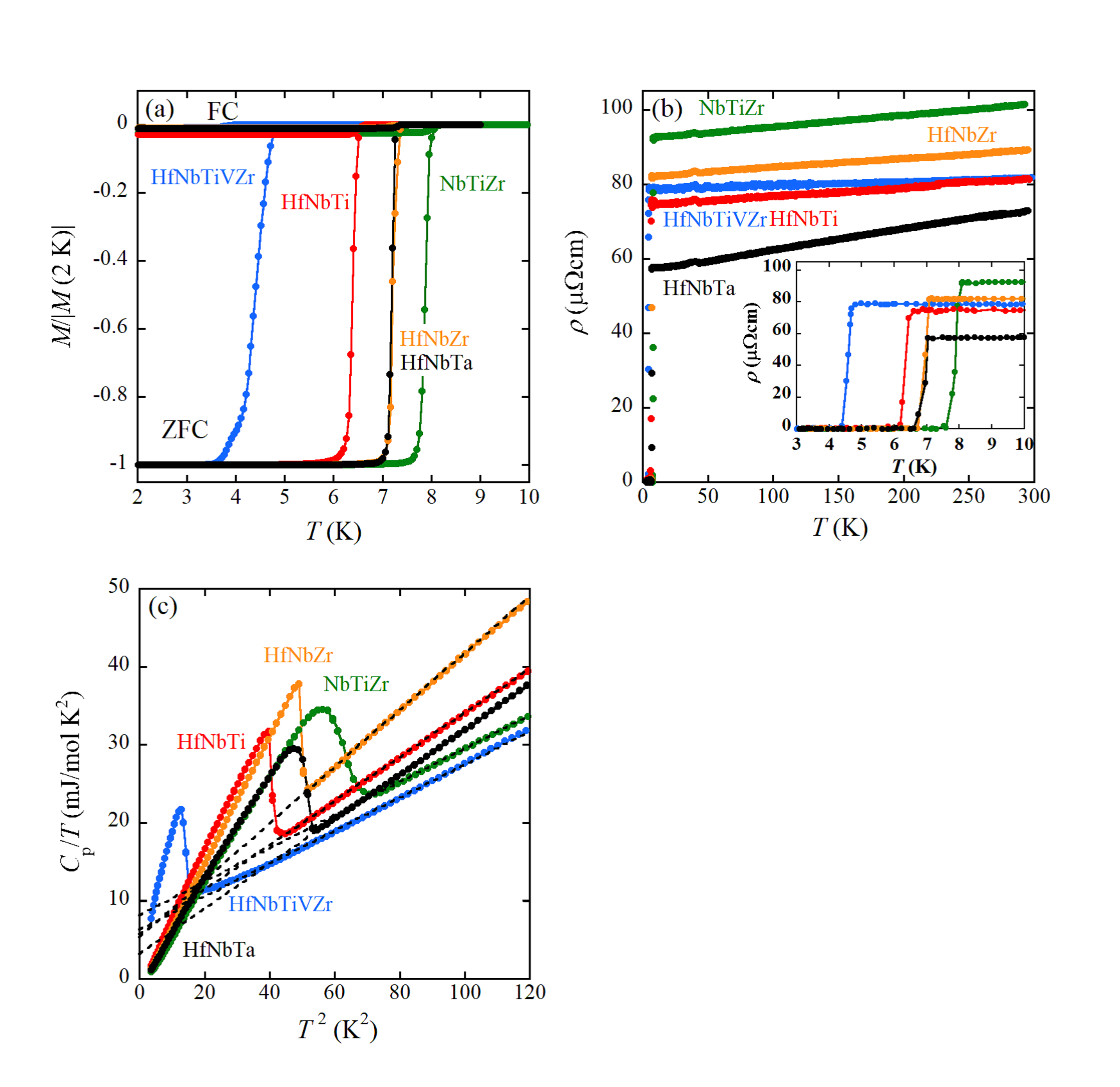}
\caption{\label{fig3} (a) Temperature-dependent magnetization of HfNbTiVZr, NbTiZr, HfNbTi, HfNbZr, and HfNbTa measured under zero-field-cooled (ZFC) and field-cooled (FC) conditions. $M$ is normalized by the absolute value of the ZFC magnetization at 2.0 K. (b) Temperature-dependent electrical resistivity of HfNbTiVZr, NbTiZr, HfNbTi, HfNbZr, and HfNbTa, with the inset displaying the low-temperature region. (c) $C_\mathrm{p}/T$ versus $T^{2}$ plots of HfNbTiVZr, NbTiZr, HfNbTi, HfNbZr, and HfNbTa. The dashed lines represent the fitting results using the relation $C_\mathrm{p}/T=\gamma +\beta T^{2}$. The($\gamma$ (mJ/mol$\cdot$K$^{2}$), $\beta$(mJ/mol$\cdot$K$^{4}$)) datasets are (6.24, 0.2115) for HfNbTiVZr, (8.12, 0.2134) for NbTiZr, (5.74, 0.2828) for HfNbTi, (5.28, 0.3640) for HfNbZr, and (3.18, 0.2877) for HfNbTa, respectively.}
\end{center}
\end{figure}

The superconducting phase transition of each sample was examined using $M$($T$) (Fig.\hspace{1mm}\ref{fig3}(a)), $\rho$($T$) (Fig.\hspace{1mm}\ref{fig3}(b)), and $C_\mathrm{p}$($T$) (Fig.\hspace{1mm}\ref{fig3}(c)). 
Figures \ref{fig3}(a) (ZFC $M$($T$)) and \ref{fig3}(b) reveal the onset of diamagnetism and zero resistivity below $T_\mathrm{c}$ for each sample, respectively. 
The normal-state resistivity data of all samples exhibit positive temperature coefficients of resistance (TCR), with values of 104 ppm/K (HfNbTiVZr), 310 ppm/K (NbTiZr), 252 ppm/K (HfNbTi), 240 ppm/K (HfNbZr), and 558 ppm/K (HfNbTa). 
Notably, the TCR of HfNbTiVZr is lower than those of the ternary alloys, probably reflecting the enhanced degree of atomic disorder associated with an increased number of constituent elements. 
The $C_\mathrm{p}$ data are crucial not only for confirming the bulk nature of the superconducting transition but also for evaluating $\theta_\mathrm{D}$. 
The sharp discontinuity observed in each sample provides clear evidence of a bulk phase transition. 
$T_\mathrm{c}$ was determined from the midpoint of the discontinuity (see also Table \ref{tab1}). 
We note that the $T_\mathrm{c}$ values of HfNbTiVZr, NbTiZr, HfNbTi, and HfNbTa have been previously reported by other research groups\cite{Dong:SM2023,Kosuth:LTP2024,Han:JALCOM2025}  based on $M$($T$) or $\rho$($T$), in agreement with the present results. 
The $C_\mathrm{p}/T$ vs. $T^{2}$ plots in Fig.\hspace{1mm}\ref{fig3}(c) above $T_\mathrm{c}$ follow a linear dependence, expressed as
\begin{equation}
C_\mathrm{p}/T=\gamma +\beta T^{2},
\end{equation} 
where $\gamma$ and $\beta$ represent the electronic specific heat coefficient and the lattice contribution, respectively. 
$\theta_\mathrm{D}$ for each sample was calculated using the relation:
\begin{equation}
\theta_\mathrm{D}=\left(\frac{12\pi^{4}RN}{5\beta}\right)^{1/3},
\end{equation}
where $R$ is the gas constant and $N$=1 is the number of atoms per formula unit. 
The resultant $\gamma$ and $\theta_\mathrm{D}$ values are summarized in Table \ref{tab1}.

We have completed the collection of $T_\mathrm{c}$ and $\theta_\mathrm{D}$ values for ten equimolar quinary HEAs and five equimolar ternary alloys (see Table \ref{tab1}). 
All $T_\mathrm{c}$ values were determined from the $C_\mathrm{p}$($T$) data. 
According to the McMillan formula\cite{McMillan:PR1968}, $T_\mathrm{c}$ is given by
\begin{equation}
T_\mathrm{c}=\frac{\theta_\mathrm{D}}{1.45}\mathrm{exp}\left[-\frac{1.04(1+\lambda_\mathrm{e-p})}{\lambda_\mathrm{e-p}-\mu^{*}(1+0.62\lambda_\mathrm{e-p})}\right].
\label{eq:mcm}
\end{equation}
Assuming a constant Coulomb pseudopotential $\mu^{*}$ = 0.13, a value commonly used for many transition-metal-based superconductors\cite{McMillan:PR1968}, $T_\mathrm{c}$ depends on both the electron–phonon coupling constant $\lambda_\mathrm{e-p}$ and $\theta_\mathrm{D}$. 
By substituting $T_\mathrm{c}$, $\theta_\mathrm{D}$, and $\mu^{*}$ = 0.13 into Eq.\hspace{1mm}(\ref{eq:mcm}), $\lambda_\mathrm{e-p}$ can be calculated. 
The $\lambda_\mathrm{e-p}$ values thus obtained for all alloys are listed in Table \ref{tab1}. 
Here, we comment on the justification for employing a fixed $\mu^{*}$ value. 
Recently, Sobota et al. performed density functional theory (DFT) calculations for HEA superconductors with varying chemical compositions and elemental combinations\cite{Sobota:ES2025}.
They evaluated $\mu^{*}$ values for all investigated HEAs using DFT and found that $\mu^{*}$ is insensitive to chemical composition and elemental combination.

\begin{table}
\caption{\label{tab1}%
Superconducting parameters and hardness of equiatomic quinary and ternary bcc HEA superconductors. Each $T_\mathrm{c}$ is determined from specific heat data. $\lambda_\mathrm{e-p}$ is recalculated using Eq.\hspace{1mm}(\ref{eq:mcm}). Hardness values are measured by our group, except for NbTaTiVZr and HfNbTaTiV.}
\begin{tabular}{ccccccc}
\br
Alloy & $T_\mathrm{c}$ (K) & $\theta_\mathrm{D}$ (K) & $\lambda_\mathrm{e-p}$ & $\gamma$ (mJ/mol$\cdot$K$^{2}$) & Hardness (HV) & Ref.\\
\mr
HfNbScTiV & 3.90 & 231(1) & 0.642(1) & 7.41(7) & 371(9) & \cite{Jangid:APL2024} \\ 
HfNbTaTiZr & 6.0 & 212 & 0.768 & 7.92 & 317(4) & \cite{Vrtnik:JALCOM2017} \\
HfNbTiVZr & 3.73(2) & 209(1) & 0.653(2) & 6.24(3) & 360(5) & this work  \\
HfMoNbTiZr & 4.1(1) & 263(1) & 0.627(4) & 5.76(2) & 398(5) & \cite{Kitagawa:JALCOM2022} \\
NbTaTiVZr & 5.13 & 215 & 0.719 & 8.60(2) & 417 & \cite{Sharma:PRM2025}  \\
HfNbTaTiV & 4.0 & 209 & 0.668 & 8.03(2) & 469 & \cite{Sharma:PRM2025}   \\
HfNbReTiZr & 5.3 & 255 & 0.686 & 5.7(1) & 526(5) & \cite{Marik:JALCOM2018}  \\
HfMoNbTaTi & 3.2 & 254 & 0.591 & 4.58 & 466(3) & \cite{Zeng:AQT2023}  \\
CuNbScTiV & 6.75(4) & 324(2) & 0.687(3) & 6.89(7) & -- & \cite{Jangid:PRM2025} \\
NbTiUVZr & 4.0 & 245(5) & 0.636(4) & 7.94(7) & -- & \cite{Nowak:AM2025} \\
HfTaTi & 4.88 & 170 & 0.772 & 5.935(6) & -- & \cite{Li:SST2024} \\
NbTiZr & 7.98(2) & 209(1) & 0.874(3) & 8.12(5) & 263(5) & this work \\
HfNbTi & 6.47(2) & 190(1) & 0.829(3) & 5.74(5) & 258(3) & this work \\
HfNbZr & 7.07(2) & 175(1) & 0.899(4) & 5.28(4) & 287(3) & this work \\
HfNbTa & 7.11(2) & 189(1) & 0.868(3) & 3.18(3) & 320(3) & this work \\
\br
\end{tabular}
\end{table}

\begin{figure}
\begin{center}
\includegraphics[width=0.6\linewidth]{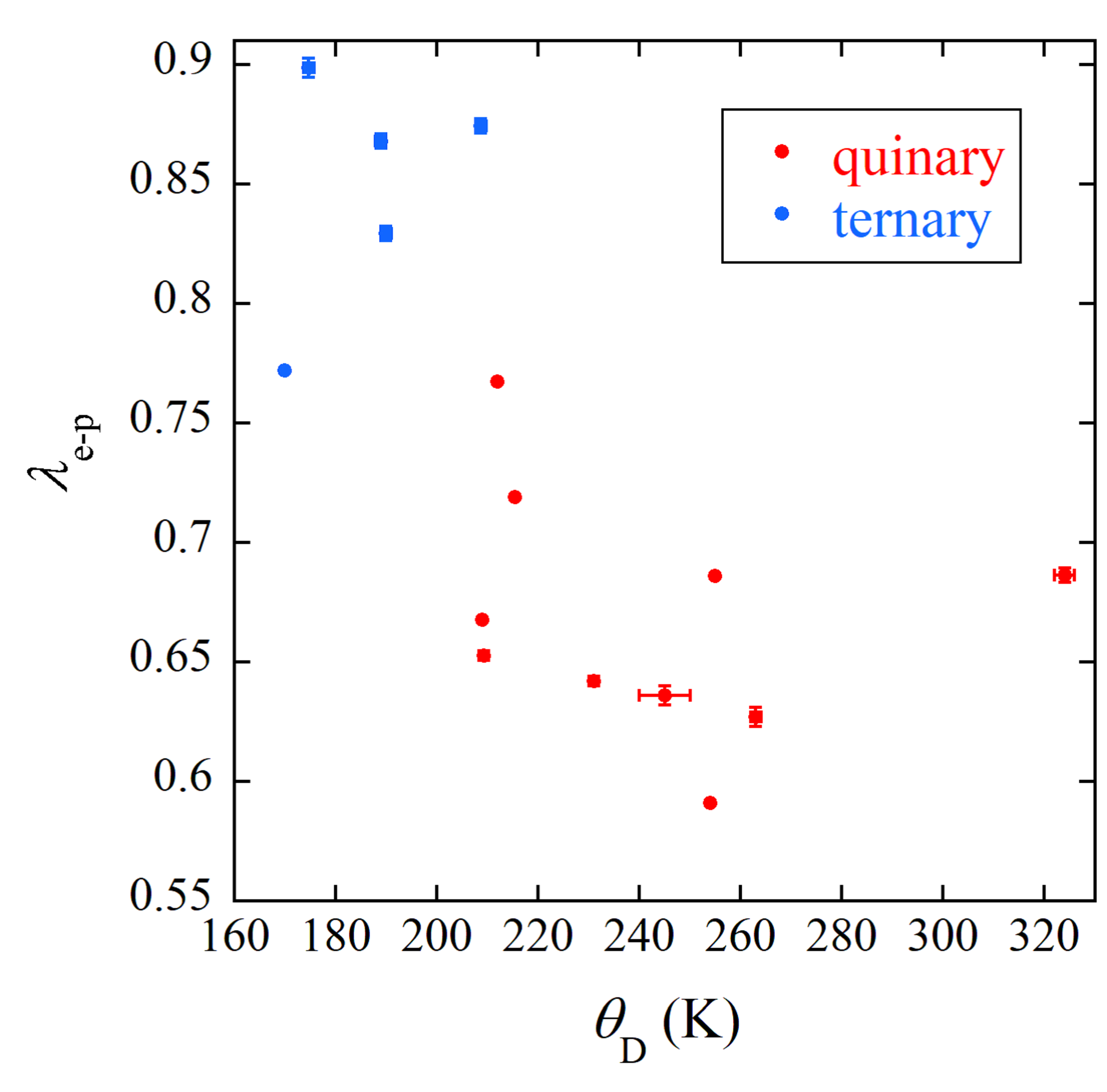}
\caption{\label{fig4} $\lambda_\mathrm{e-p}$ versus $\theta_{D}$ plot for equimolar quinary and ternary alloys.}
\end{center}
\end{figure}

{\footnotesize
\begin{longtable}{ccccccc}
\caption{\label{tab2}%
Superconducting parameters of HEA superconductors comprising two to six constituent elements. Each $T_\mathrm{c}$ is determined from specific heat data. $\lambda_\mathrm{e-p}$ is recalculated using Eq.\hspace{1mm}(\ref{eq:mcm}). The datasets for equiatomic quinary and ternary alloys are provided in Table \ref{tab1}.}\\
\hline
Alloy & $T_\mathrm{c}$ (K) & $\theta_\mathrm{D}$ (K) & $\lambda_\mathrm{e-p}$ & $\gamma$ (mJ/mol$\cdot$K$^{2}$) & Hardness (HV) & Ref.\\
\hline
\endfirsthead
\hline
Alloy & $T_\mathrm{c}$ (K) & $\theta_\mathrm{D}$ (K) & $\lambda_\mathrm{e-p}$ & $\gamma$ (mJ/mol$\cdot$K$^{2}$) & Hardness (HV) & Ref.\\
\hline
\endhead
\hline
\endfoot
\hline
\endlastfoot

(TaNbV)$_{0.67}$(ZrHfTi)$_{0.33}$ & 4.11 & 236 & 0.648 & 7.97 & -- & \cite{Rohr:PNAS2016} \\ 
Ta$_{34}$Nb$_{33}$Hf$_{8}$Zr$_{14}$Ti$_{11}$ & 7.3 & 243(5) & 0.786(6) & 8.3 & -- & \cite{Kozelj:PRL2014} \\
(TaNb)$_{0.67}$(HfZrTi)$_{0.33}$ & 7.7 & 225(2) & 0.831(3) & 7.97(5) & -- & \cite{Rohr:PNAS2016} \\
(TaNb)$_{0.16}$(HfZrTi)$_{0.84}$ & 4.59 & 184(2) & 0.731(3) & 6.45(8) & -- & \cite{Rohr:PNAS2016} \\
Ta$_{1/6}$Nb$_{2/6}$Hf$_{1/6}$Zr$_{1/6}$Ti$_{1/6}$ & 7.85 & 192 & 0.903 & 7.45 & -- & \cite{Kim:AM2020} \\
(TaNb)$_{0.31}$(HfUTi)$_{0.69}$ & 2.5 & 193 & 0.595 & 7.1 & -- & \cite{Nelson:SR2020} \\
(NbTa)$_{0.67}$(MoHfW)$_{0.33}$ & 4.3(1) & 256(1) & 0.641(2) & 4.8(1) & -- & \cite{Sobota:PRB2022} \\
(NbTi)$_{0.67}$(MoHfV)$_{0.33}$ & 4.8(2) & 315(5) & 0.62(1) & 7.4(1) & -- & \cite{Nowak:MMTA2024} \\
Ti$_{35}$Hf$_{25}$Nb$_{25}$Ta$_{5}$Re$_{10}$ & 3.95(2) & 233(1) & 0.643(2) & 5.33(3) & 439(7) & \cite{Hattori:JAMS2023} \\
Ti$_{30}$Hf$_{20}$Nb$_{35}$Ta$_{5}$Re$_{10}$ & 4.38(2) & 241(1) & 0.657(2) & 5.61(4) & 428(5) & \cite{Hattori:JAMS2023} \\
Ti$_{25}$Hf$_{15}$Nb$_{35}$Ta$_{15}$Re$_{10}$ & 4.10(2) & 250(1) & 0.637(2) & 5.52(4) & 446(9) & \cite{Hattori:JAMS2023} \\
Ti$_{20}$Hf$_{10}$Nb$_{35}$Ta$_{25}$Re$_{10}$ & 3.62(2) & 260(1) & 0.607(2) & 5.30(3) & 460(10) & \cite{Hattori:JAMS2023} \\
Ti$_{15}$Hf$_{5}$Nb$_{35}$Ta$_{35}$Re$_{10}$ & 3.25(2) & 261(1) & 0.589(2) & 5.37(3) & 466(6) & \cite{Hattori:JAMS2023} \\
Ti$_{0.5}$(ZrNbHfTa)$_{0.5}$ & 6.0(1) & 184(5) & 0.81(2) & 7.9(4) & -- & \cite{Sobota:AM2025} \\
Ti$_{0.5}$(VNbHfTa)$_{0.5}$ & 5.9(1) & 205(1) & 0.77(1) & 7.9(7) & -- & \cite{Sobota:AM2025} \\
(Ti$_{1/4}$Hf$_{1/4}$Nb$_{1/4}$Ta$_{1/4}$)$_{0.986}$Ni$_{0.014}$ & 6.55 & 204 & 0.808 & 4.62 & -- &           \cite{Huang:SST2025} \\
(Ti$_{1/4}$Hf$_{1/4}$Nb$_{1/4}$Ta$_{1/4}$)$_{0.973}$Ni$_{0.027}$ & 6.70 & 220 & 0.791 & 4.66 & -- &           \cite{Huang:SST2025} \\
(Ti$_{1/4}$Hf$_{1/4}$Nb$_{1/4}$Ta$_{1/4}$)$_{0.958}$Ni$_{0.042}$ & 6.93 & 206 & 0.825 & 4.20 & -- &           \cite{Huang:SST2025} \\
(Ti$_{1/4}$Hf$_{1/4}$Nb$_{1/4}$Ta$_{1/4}$)$_{0.941}$Ni$_{0.059}$ & 6.96 & 211 & 0.818 & 4.71 & -- &           \cite{Huang:SST2025} \\
(Ti$_{1/4}$Hf$_{1/4}$Nb$_{1/4}$Ta$_{1/4}$)$_{0.923}$Ni$_{0.077}$ & 7.36 & 211 & 0.838 & 4.3 & -- &           \cite{Huang:SST2025} \\
Hf$_{35}$Nb$_{10}$Ta$_{35}$Mo$_{10}$W$_{10}$ & 4.06(2) & 177.3(8) & 0.709(3) & 4.23(1) & -- & \cite{Krishnan:SST2025} \\
Hf$_{35}$Nb$_{22}$Ta$_{23}$Mo$_{10}$W$_{10}$ & 4.82(2) & 208.8(1) & 0.711(1) & 6.18(9) & -- & \cite{Krishnan:SST2025} \\
Hf$_{35}$Nb$_{35}$Ta$_{10}$Mo$_{10}$W$_{10}$ & 5.91(2) & 212.1(1) & 0.763(1) & 7.49(4) & -- & \cite{Krishnan:SST2025} \\
Mo$_{35}$W$_{35}$Re$_{15}$Ru$_{10}$Pd$_{5}$ & 3.98 & 365 & 0.570 & 3.15 & -- & \cite{Xu:SM2024} \\
Re$_{23}$Os$_{5}$W$_{33}$Mo$_{34}$Ta$_{5}$ & 4.42 & 327 & 0.602 & 4.96(2) & -- & \cite{Krishnan:SST2025-2} \\
Hf$_{5}$Nb$_{45}$Sc$_{10}$Ti$_{5}$Zr$_{35}$ & 7.73(2) & 207(1) & 0.865(3) & 7.62(3) & -- & \cite{Kubo:JALCOM2025} \\
Ti$_{15}$Zr$_{15}$Nb$_{35}$Ta$_{35}$ & 8.00 & 191(3) & 0.914(7) & 9.3(1) & -- & \cite{Yuan:FM2018} \\	
TiVNbTa & 4.38 & 166.54	& 0.746 & 3.023(3) & -- & \cite{Li:JPC2023} \\	
TiHfNbTa & 6.30 & 199.9 & 0.802 & 4.698(5) & -- & \cite{Zeng:SCPMA:2023} \\	
Nb$_{2/5}$Hf$_{1/5}$Zr$_{1/5}$Ti$_{1/5}$ & 7.99 & 193 & 0.909 &  5.937 & -- & \cite{Hidayati:AM2023} \\	
(Ti$_{1/3}$Hf$_{1/3}$Ta$_{1/3}$)$_{0.965}$Nb$_{0.035}$ & 5.61(1) & 180.6 & 0.797(1) & 5.713(2) & -- & \cite{Li:SST2024} \\	
(Ti$_{1/3}$Hf$_{1/3}$Ta$_{1/3}$)$_{0.875}$Nb$_{0.125}$ & 5.82(2) & 181.5 & 0.808(1) & 6.043(2) & -- & \cite{Li:SST2024} \\	
(Ti$_{1/3}$Hf$_{1/3}$Ta$_{1/3}$)$_{0.67}$Nb$_{0.33}$ & 7.06(6) & 198.1 & 0.846(3) & 4.838(6) & -- & \cite{Li:SST2024} \\	
(Ti$_{1/3}$Hf$_{1/3}$Ta$_{1/3}$)$_{0.56}$Nb$_{0.44}$ & 7.41(6) & 205.8 & 0.850(3) & 5.691(4) & -- & \cite{Li:SST2024} \\	
(Ti$_{1/3}$Hf$_{1/3}$Ta$_{1/3}$)$_{0.43}$Nb$_{0.57}$ & 8.19(2) & 221.4 & 0.861(2) & 6.161(3) & -- & \cite{Li:SST2024} \\	
(Ti$_{1/3}$Hf$_{1/3}$Ta$_{1/3}$)$_{0.25}$Nb$_{0.75}$ & 8.82(5) & 243.9 & 0.852(3) & 6.840(2) & -- & \cite{Li:SST2024} \\	
(Ti$_{1/3}$Hf$_{1/3}$Ta$_{1/3}$)$_{0.1}$Nb$_{0.9}$ & 9.11(1) & 253.0 & 0.850(1) & 8.243(3) & -- & \cite{Li:SST2024} \\	
(TiV)$_{0.5}$Nb$_{0.4}$Ta$_{0.1}$ & 4.50(5) & 255.8(1) & 0.650(2) &	10.10(2) & -- & \cite{Sharma:MTC2025} \\	
HfNbZrTi & 6.00 & 290.1 & 0.685 & 2.91 & -- & \cite{Sang:SST2025} \\	
TaNbHfZr & 6.50 & 352(6) & 0.660(4) & 6.6376(2) & -- & \cite{Krishnan:JALCOM2025}\\	
(NbHf)$_{0.75}$(TiZr)$_{0.25}$ & 7.15(5) & 164(1) & 0.933(6) & 7.8(3) & -- & \cite{Idczak:PRB2025}\\	
(NbZr)$_{0.75}$(TiHf)$_{0.25}$ & 8.17(5) & 170(1) & 0.984(7) & 10.8(9) & -- & \cite{Idczak:PRB2025}\\	
(NbTi)$_{0.75}$(ZrHf)$_{0.25}$ & 7.76(5) & 201(1) & 0.879(4) & 8.8(3) & -- & \cite{Idczak:PRB2025}\\	
Nb$_{0.25}$Ta$_{0.25}$Ti$_{0.25}$Zr$_{0.25}$ & 8.0 & 206(1) & 0.881(2) & 5.51(5) & -- & \cite{Sharma:PRB} \\
Zr$_{0.02}$Nb$_{0.96}$Mo$_{0.02}$ & 8.93(1) & 263(2) & 0.828(3) & 7.69(2) & -- & \cite{Ishikawa:JLTP1975} \\
Zr$_{0.05}$Nb$_{0.9}$Mo$_{0.05}$ & 8.50(1) & 265(2) & 0.808(3) & 7.53(2) & -- & \cite{Ishikawa:JLTP1975} \\
Zr$_{0.1}$Nb$_{0.8}$Mo$_{0.1}$ & 7.78(2) & 258(2) & 0.787(3) & 7.06(2) & -- & \cite{Ishikawa:JLTP1975} \\
V$_{0.095}$Nb$_{0.298}$Ta$_{0.607}$ & 4.47(2) & 259.2(4) & 0.646(1) & 6.52(3) & -- & \cite{Wang:JLTP1978} \\
V$_{0.212}$Nb$_{0.42}$Ta$_{0.368}$ & 4.41(4) & 267.1(8) & 0.638(2) & 7.00(6) & -- & \cite{Wang:JLTP1978} \\
V$_{0.251}$Nb$_{0.578}$Ta$_{0.171}$ & 4.67(7) & 267.9(8) & 0.648(4) & 7.54(7) & -- & \cite{Wang:JLTP1978} \\
V$_{4}$Ti$_{2}$W & 5.0 & 171 & 0.778 & 6.08 & -- & \cite{Li:MTC2024} \\
Nb$_{2}$TiW & 4.2 & 304 & 0.606 & 4.681 & -- & \cite{Li:CPL2025} \\
Nb$_{2}$TiMo & 2.8 & 322 & 0.540 & 5.314 & -- & \cite{Li:CPL2025} \\
V$_{0.85}$Nb$_{0.15}$ &	4.30(2) & 364(10) &	0.581(5) & 9.2(2) & -- & \cite{Corsan:PSS1970}\\
V$_{0.7}$Nb$_{0.3}$ & 4.03(2) & 326(10) & 0.588(5) & 8.7(2) & -- & \cite{Corsan:PSS1970} \\
V$_{0.5}$Nb$_{0.5}$ & 4.25(2) & 298(9) & 0.611(6) &	8.3(2) & -- & \cite{Corsan:PSS1970}\\
V$_{0.31}$Nb$_{0.69}$ &	4.90(2) & 285(9) & 0.646(6) & 8.0(2) & -- & \cite{Corsan:PSS1970}\\
V$_{0.2}$Nb$_{0.8}$ & 5.76(3) & 275(8) & 0.688(7) & 7.9(2) & -- & \cite{Corsan:PSS1970}\\
V$_{0.1}$Nb$_{0.9}$ & 6.95(3) & 266(8) & 0.74(1) & 7.8(2) & -- & \cite{Corsan:PSS1970} \\
V$_{0.9}$Ta$_{0.1}$ & 4.47(2) & 341(10) & 0.597(4) & 8.8(3) & -- & \cite{Corsan:PSS1970}\\
V$_{0.76}$Ta$_{0.24}$ &	3.58(2) & 300(9) & 0.583(4) & 8.5(3) & -- & \cite{Corsan:PSS1970}\\
V$_{0.5}$Ta$_{0.5}$ & 2.73(1) & 273(8) & 0.558(4) & 7.6(2) & -- & \cite{Corsan:PSS1970}\\
V$_{0.3}$Ta$_{0.7}$ & 3.00(2) &	256(8) & 0.580(6) &	6.4(2) & -- & \cite{Corsan:PSS1970}\\
V$_{0.1}$Ta$_{0.9}$ & 3.58(2) & 250(8) & 0.612(6) &	5.7(2) & -- & \cite{Corsan:PSS1970}\\
Nb$_{0.8}$Ta$_{0.2}$ & 7.42(4) & 257(8) & 0.77(1) &	7.3(2) & -- & \cite{Corsan:PSS1970}\\
Nb$_{0.7}$Ta$_{0.3}$ & 6.80(3) & 263(8) & 0.741(9) & 6.8(2) & -- & \cite{Corsan:PSS1970}\\
Nb$_{0.5}$Ta$_{0.5}$ & 5.93(3) & 260(8) & 0.708(8) & 6.5(2) & -- & \cite{Corsan:PSS1970}\\
Nb$_{0.3}$Ta$_{0.7}$ & 5.15(3) & 260(8) & 0.675(8) & 6.2(2) & -- & \cite{Corsan:PSS1970}\\
Nb$_{0.15}$Ta$_{0.85}$ & 4.58(2) & 255(8) &	0.654(8) & 6.0(2) & -- & \cite{Corsan:PSS1970}\\
Ti$_{0.87}$Mo$_{0.13}$ & 3.96 &	252.4 &	0.628 &	6.81 & -- & \cite{Sinha:JPCS1968}\\	
Ti$_{0.85}$Mo$_{0.15}$ & 4.06 &	276.2 &	0.617 &	7.23 & -- & \cite{Sinha:JPCS1968}\\	
Ti$_{0.8}$Mo$_{0.2}$ & 4.00 & 299.0 & 0.601 & 7.40 & -- & \cite{Sinha:JPCS1968}\\	
Ti$_{0.7}$Mo$_{0.3}$ & 3.61 & 300.0 & 0.584 & 6.67 & -- & \cite{Sinha:JPCS1968}\\	
Ti$_{0.6}$Mo$_{0.4}$ & 2.60 & 324.0 & 0.530 & 5.95 & -- & \cite{Sinha:JPCS1968}\\	
Ta$_{0.975}$Re$_{0.025}$ & 3.458 & 261 & 0.599 & 5.70 & -- & \cite{Mamiya:JPSJ1970}\\	
Ta$_{0.95}$Re$_{0.05}$ & 2.77 & 277 & 0.558 & 5.12 & -- & \cite{Mamiya:JPSJ1970}\\	
Ta$_{0.925}$Re$_{0.075}$ & 2.08 & 285 & 0.519 & 5.05 & -- & \cite{Mamiya:JPSJ1970}\\	
Ta$_{0.9}$Re$_{0.1}$ & 1.49 & 296 & 0.482 & 4.10 & -- & \cite{Mamiya:JPSJ1970}\\	
Ta$_{0.85}$Re$_{0.15}$ & 0.75 & 307 & 0.428 & 3.75 & -- & \cite{Mamiya:JPSJ1970}\\	
Ta$_{0.8}$Re$_{0.2}$ & 0.21 & 317 & 0.363 & 3.00 & -- & \cite{Mamiya:JPSJ1970}\\	
Ti$_{0.7}$V$_{0.3}$ & 6.14 & 244 & 0.734 & 10.0 & -- & \cite{Cheng:PR1962}\\	
Ti$_{0.5}$V$_{0.5}$ & 7.30 & 262 & 0.763 & 10.8 & -- & \cite{Cheng:PR1962}\\	
Ti$_{0.25}$V$_{0.75}$ & 7.16 & 279 & 0.739 & 10.6 & -- & \cite{Cheng:PR1962}\\	
Ti$_{0.15}$V$_{0.85}$ & 7.02 & 283 & 0.730 & 10.3 & -- & \cite{Cheng:PR1962}\\	
Nb$_{0.4}$Zr$_{0.6}$ & 8.8 & 190 & 0.964 & 15.9 & -- & \cite{Morin:PR1963}\\	
Nb$_{0.75}$Zr$_{0.25}$ & 10.78 & 255 & 0.919 & 8.96 & -- & \cite{Junod:JLTP1986}\\	
Nb$_{0.9}$Zr$_{0.1}$ & 10.5 & 225 & 0.968 & 11.7 & -- & \cite{Morin:PR1963}\\	
Nb$_{0.9}$Mo$_{0.1}$ & 5.3 & 275 & 0.669 & 7.5 & -- & \cite{Morin:PR1963}\\	
Nb$_{0.75}$Mo$_{0.25}$ & 3.4 & 305 & 0.573 & 5.6 & -- & \cite{Morin:PR1963}\\	
Nb$_{0.62}$Mo$_{0.38}$ & 0.76 & 325 & 0.425 & 3.5 & -- & \cite{Morin:PR1963}\\	
Nb$_{0.6}$Mo$_{0.4}$ & 0.50 & 340 & 0.399 & 3.0 & -- & \cite{Morin:PR1963}\\	
Nb$_{0.58}$Mo$_{0.42}$ & 0.31 & 340 & 0.376 & 2.7 & -- & \cite{Morin:PR1963}\\	
Mo$_{0.95}$Re$_{0.05}$ & 1.5 & 450 & 0.449 & 2.2 & -- & \cite{Morin:PR1963}\\	
Mo$_{0.9}$Re$_{0.1}$ & 2.9 & 440 & 0.508 & 2.6 & -- & \cite{Morin:PR1963}\\	
Mo$_{0.8}$Re$_{0.2}$ & 8.5 & 420 & 0.680 & 3.8 & -- & \cite{Morin:PR1963}\\	
Mo$_{0.75}$Re$_{0.25}$ & 10.5 & 405 & 0.742 & 4.0 & -- & \cite{Morin:PR1963} \\	
Mo$_{0.7}$Re$_{0.3}$ & 10.8 & 395 & 0.757 & 4.1 & -- & \cite{Morin:PR1963}\\	
Mo$_{0.6}$Re$_{0.4}$ & 12.6 & 340 & 0.862 & 4.4 & -- & \cite{Morin:PR1963}\\	
Mo$_{0.5}$Re$_{0.5}$ & 11.5 & 320 & 0.850 &	4.4 & -- & \cite{Morin:PR1963}\\	
V$_{0.95}$Mo$_{0.05}$ & 3.73 & 430 & 0.539 & 8.85 & -- & \cite{Kuentzler1986}\\	
V$_{0.9}$Mo$_{0.1}$ & 2.50 & 417 & 0.499 & 7.55 & -- & \cite{Kuentzler1986}\\
Ta$_{0.95}$Zr$_{0.05}$ & 5.31 & 263(13) & 0.68(1) & 6.6(1) & -- & \cite{Klimczuk:PRM2023}\\
Ta$_{0.9}$Zr$_{0.1}$ & 6.3(5) & 235(10) & 0.75(4) & 7.2(1) & -- & \cite{Klimczuk:PRM2023}\\
Ta$_{0.85}$Zr$_{0.15}$ & 6.9(5) & 230(10) & 0.78(4) & 7.4(1) & -- & \cite{Klimczuk:PRM2023}\\
Ta$_{0.95}$Hf$_{0.05}$ & 5.21 & 260(10) & 0.678(7) & 6.6(1) & -- & \cite{Klimczuk:PRM2023}\\
Ta$_{0.9}$Hf$_{0.1}$ & 5.8(5) & 250(10) & 0.71(3) & 7.5(1) & -- & \cite{Klimczuk:PRM2023}\\
Ta$_{0.85}$Hf$_{0.15}$ & 6.4(5) & 255(10) &	0.73(3) & 7.6(1) & -- & \cite{Klimczuk:PRM2023}\\
Ta$_{0.8}$Hf$_{0.2}$ & 6.6 & 256.2 & 0.740 & 8.0 & -- & \cite{Meena:PRM2023}\\	
Ta$_{0.7}$Hf$_{0.3}$ & 6.6(5) & 190(10) & 0.84(4) & 7.3(3) & -- & \cite{Klimczuk:PRM2023}\\
Ta$_{0.6}$Hf$_{0.4}$ & 6.7 & 222.3 & 0.787 & 9.3 & -- & \cite{Meena:PRM2023}\\	
Ta$_{0.5}$Hf$_{0.5}$ & 6.1 & 151.6 & 0.896 & 8.9 & -- & \cite{Meena:PRM2023}\\	
Ta$_{0.4}$Hf$_{0.6}$ & 5.6 & 158.7 & 0.843 & 6.6 & -- & \cite{Meena:PRM2023}\\
Ta$_{0.3}$Hf$_{0.7}$ & 4.7(5) & 155(10) & 0.79(4) & 5.5(1) & -- & \cite{Klimczuk:PRM2023} \\
Nb$_{0.9}$Ti$_{0.1}$ & 8.90	& 261 & 0.830 & 10.25 & -- & \cite{Suk:SPJ1971,Collings:book}\\	
Nb$_{0.75}$Ti$_{0.25}$ & 9.72 & 265 & 0.858 & 10.22 & -- & \cite{Suk:SPJ1971,Collings:book}\\	
Nb$_{0.5}$Ti$_{0.5}$ & 9.32 & 236 & 0.888 & 10.66 & -- & \cite{Suk:SPJ1971,Collings:book}\\	
Nb$_{0.25}$Ti$_{0.75}$ & 6.09 & 252 & 0.723 & 8.45 & -- & \cite{Suk:SPJ1971,Collings:book}\\	
Nb$_{0.2}$Ti$_{0.8}$ & 6.95 & 228 & 0.791 & 9.13 &-- & \cite{Suk:SPJ1971,Collings:book}\\
\end{longtable}
}

\begin{figure}
\begin{center}
\includegraphics[width=1.0\linewidth]{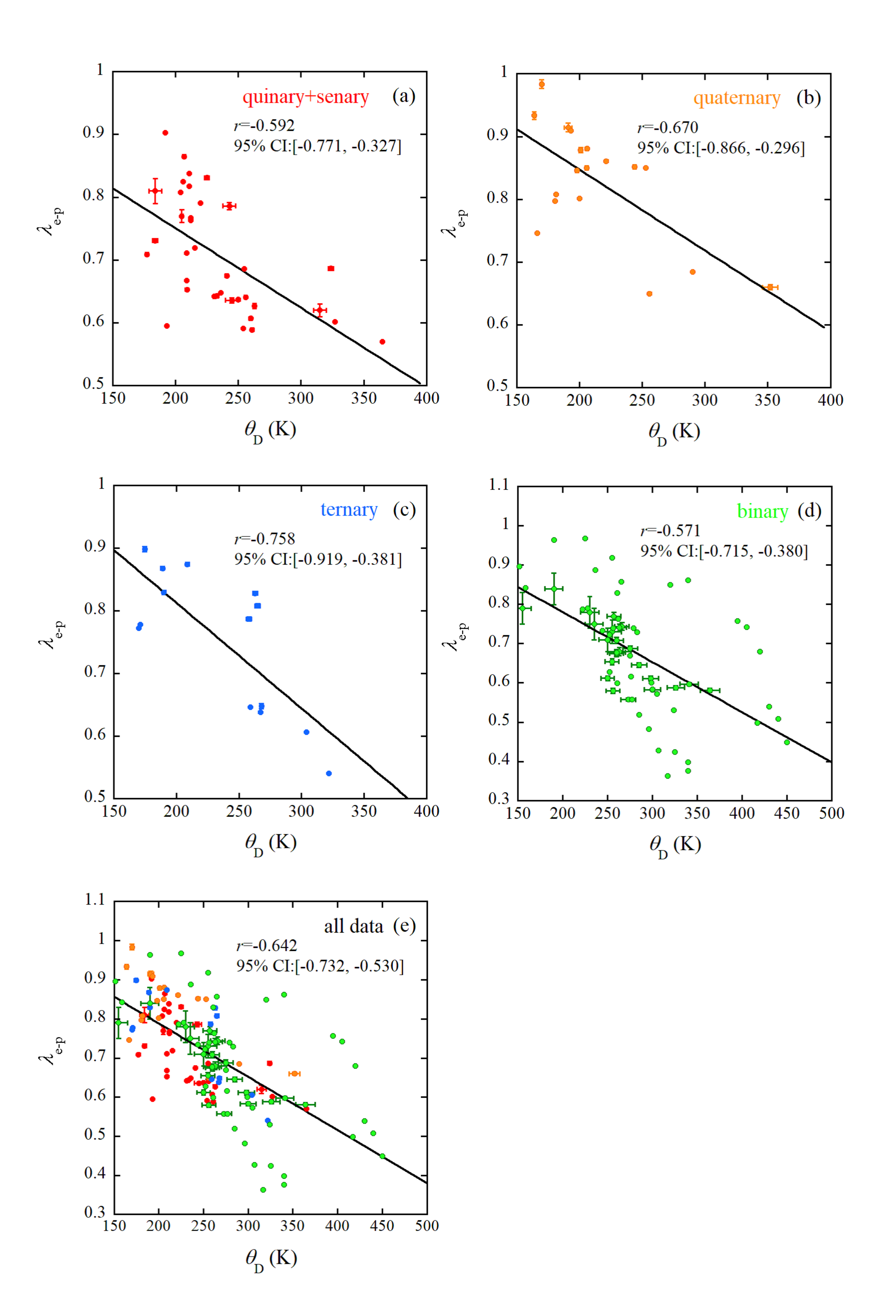}
\caption{\label{fig5} $\lambda_\mathrm{e-p}$ versus $\theta_{D}$ plot for (a) senary and quinary bcc alloys, (b) quaternary bcc alloys, (c) ternary bcc alloys, (d) binary bcc alloys, and (e) all bcc alloys.}
\end{center}
\end{figure}

\begin{figure}
\begin{center}
\includegraphics[width=1.0\linewidth]{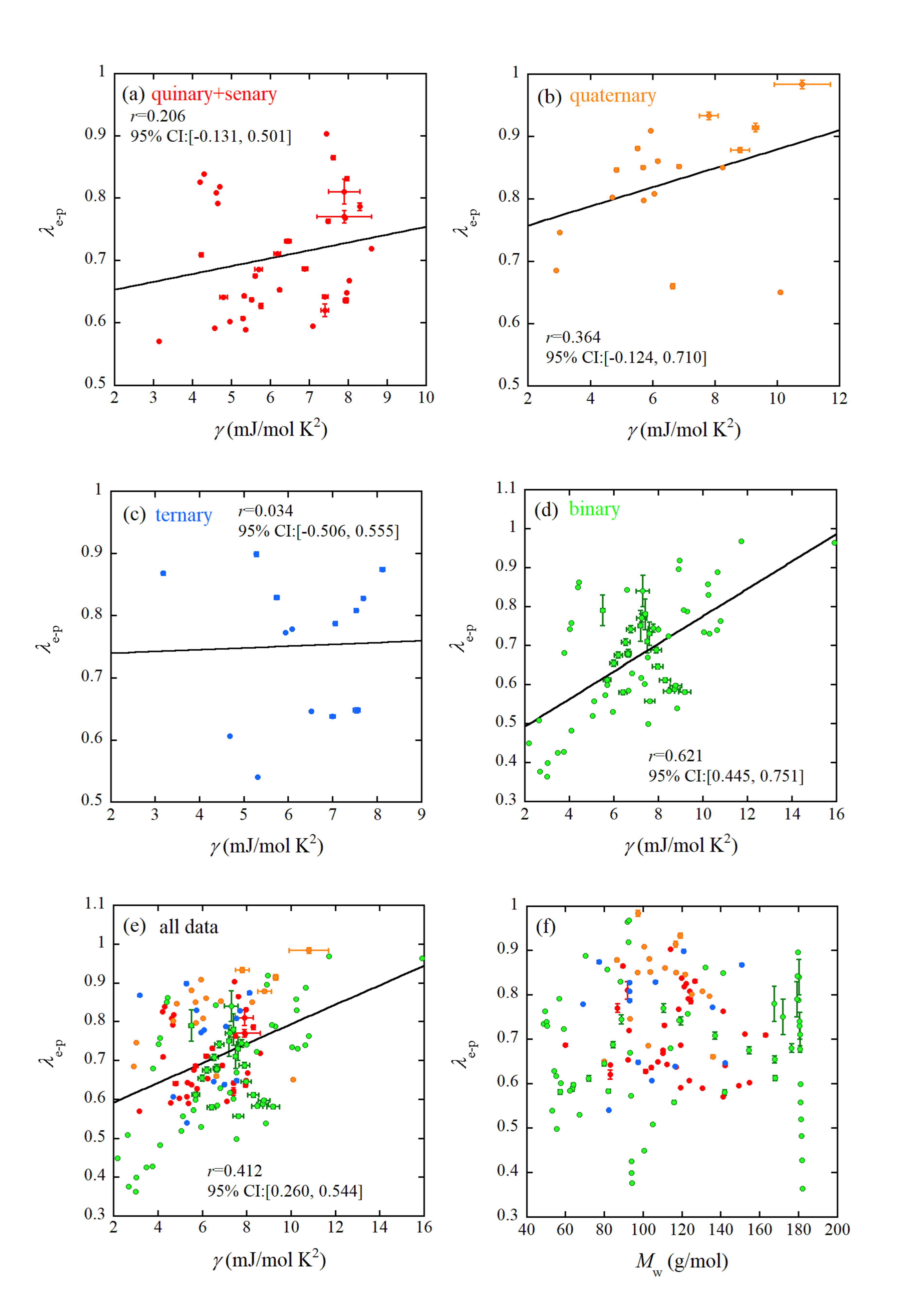}
\caption{\label{fig6} $\lambda_\mathrm{e-p}$ versus $\gamma$ plot for (a) senary and quinary bcc alloys, (b) quaternary bcc alloys, (c) ternary bcc alloys, (d) binary bcc alloys, and (e) all bcc alloys. (f) $\lambda_\mathrm{e-p}$ versus $M_\mathrm{w}$ plot for all alloys.}
\end{center}
\end{figure}

The principal aim of this study is to verify our hypothesis that the uncertainty principle, relating energy uncertainty to phonon lifetime, influences $T_\mathrm{c}$ in quinary HEA superconductors. 
This hypothesis proposes a negative correlation between $\lambda_\mathrm{e-p}$ and $\theta_\mathrm{D}$.
Figure \ref{fig4} displays $\lambda_\mathrm{e-p}$ versus $\theta_\mathrm{D}$ for all alloys (red circles: quinary; blue circles: ternary). 
In our previous work, analysis of four quinary alloys (HfNbTaTiZr, HfNbReTiZr, HfNbTaTiV, and HfMoNbTiZr) revealed that increasing $\theta_\mathrm{D}$ tends to reduce $\lambda_\mathrm{e-p}$. 
This trend is further supported by additional quinary HEAs—HfNbScTiV, HfNbTiVZr, NbTaTiVZr, HfMoNbTaTi, CuNbScTiV, and NbTiUVZr. 
Within our framework, the negative correlation between $\lambda_\mathrm{e-p}$ and $\theta_\mathrm{D}$ is attributed to the uncertainty principle.
The high degree of atomic disorder in equiatomic five-component alloys causes severe phonon broadening in the dispersion relation\cite{Kormann:npjCM2017}.
A higher $\theta_\mathrm{D}$ contributes to a greater energy uncertainty, which, according to the uncertainty principle, shortens phonon lifetimes, thereby reducing $\lambda_\mathrm{e-p}$ and explaining the negative correlation. 
In contrast, this mechanism is expected to be less effective in equiatomic ternary alloys, where atomic disorder is significantly reduced. 
Thus, a markedly different correlation might be anticipated. 
However, the merged dataset in Fig.\hspace{1mm}\ref{fig4}, including ternary alloys (blue circles), suggests a universal $\lambda_\mathrm{e-p}$-$\theta_\mathrm{D}$ relationship, rather than a distinct separation between the two systems. 

The indication of a possible universal $\lambda_\mathrm{e-p}$-$\theta_\mathrm{D}$ relationship motivated us to compile all reported ($T_\mathrm{c}$, $\theta_\mathrm{D}$, $\gamma$) datasets for bcc alloys spanning binary to senary systems, as listed in Table \ref{tab2}. 
Because only a single dataset exists for the senary system and no single-phase bcc alloy containing seven or more constituent elements has been reported, we included the senary dataset together with those of the quinary alloys. 
Figures \ref{fig5}(a)-(d) present the $\lambda_\mathrm{e-p}$-$\theta_\mathrm{D}$ relationships for each system, along with regression analyses. 
In each panel, the Pearson correlation coefficient (Pearson's $r$) and the 95 \% confidence interval (CI) are reported. 
The analyses indicate that all systems exhibit a common negative correlation coefficient with high reliability. 
The complete dataset, analyzed in Fig. \hspace{1mm}\ref{fig5}(e), yields $r$=-0.642 with a 95 \% CI of [-0.732, -0.530]. 
These regression analyses support the conclusion that bcc alloys obey a universal $\lambda_\mathrm{e-p}$-$\theta_\mathrm{D}$ relationship with a negative correlation, irrespective of the degree of atomic disorder.

In the McMillan formalism, $\lambda_\mathrm{e-p}$ is defined as
\begin{equation}
\lambda_\mathrm{e-p}=\frac{N(E_\mathrm{F})\langle I^{2}\rangle}{M_\mathrm{w}\langle\omega^{2}\rangle},
\label{eq:e-p}
\end{equation}
where $N(E_\mathrm{F})$ is the density of states at the Fermi level $E_\mathrm{F}$, $M_\mathrm{w}$ is the molecular weight, and $\langle I^{2}\rangle$ and $\langle\omega^{2}\rangle$ denote the averaged square of electron–phonon matrix element and the averaged square of phonon frequency, respectively\cite{McMillan:PR1968}. 
In this context, $N(E_\mathrm{F})$ and $\sqrt{\langle\omega^{2}\rangle}$ correspond to $\gamma$ and $\theta_{D}$\cite{Dynes:PRB1970,Gabovich:LTP2025}, respectively. 
Since the evaluation of $\langle I^{2}\rangle$ is not straightforward, this term is tentatively excluded from the following discussion. 
Thus, $\lambda_\mathrm{e-p}$ is determined primarily by $\gamma$, $M_\mathrm{w}$, and $\theta_{D}$. 
Indeed, the negative correlation between $\lambda_\mathrm{e-p}$ and $\theta_{D}$ shown in Fig.\hspace{1mm}\ref{fig5}(e) is consistent with Eq.\hspace{1mm}(\ref{eq:e-p}). 
To further examine these dependencies, we plotted $\lambda_\mathrm{e-p}$ versus $\gamma$ and $\lambda_\mathrm{e-p}$ versus $M_\mathrm{w}$ in Figs.\hspace{1mm}\ref{fig6}(a)-(f).
Although no apparent correlation is observed in the $\lambda_\mathrm{e-p}$ versus $M_\mathrm{w}$ plot (Fig.\hspace{1mm}\ref{fig6}(f)), a positive correlation ($r$=0.412 with a 95 \% CI of [0.260, 0.544]) is obtained for the $\lambda_\mathrm{e-p}$ versus $\gamma$ plot encompassing all systems (Fig.\hspace{1mm}\ref{fig6}(e)). 
However, this trend appears to be driven primarily by the binary system (Fig.\hspace{1mm}\ref{fig6}(d)), which exhibits a stronger positive correlation. 
The binary system benefits from extensive investigation, and the large number of datasets ensures reliable regression analysis. 
This positive correlation is consistent with Eq.\hspace{1mm}(\ref{eq:e-p}) and also aligns with the Matthias rule, which asserts that $T_\mathrm{c}$ increases with increasing $N(E_\mathrm{F})$ ($\gamma$) through enhancement of $\lambda_\mathrm{e-p}$. 
Although the datasets for ternary or quaternary systems (Figs.\hspace{1mm}\ref{fig6}(b) and (c)) remain limited, recent rapid progress in the study of HEA superconductors has yielded a comparatively rich dataset for quinary systems. 
The $r$ values and 95 \% CI noted in Fig.\hspace{1mm}\ref{fig6}(a) for quinary alloys indicate a weak positive correlation with limited reliability, differing from the binary system. 
Therefore, at the present stage, we conclude that no universal $\lambda_\mathrm{e-p}$ versus $\gamma$ relationship holds across binary to senary systems. 
Our analysis suggests that $\theta_{D}$ plays the dominant role in determining $\lambda_\mathrm{e-p}$, regardless of the number of alloying elements—an insight that would be valuable for designing bcc superconducting alloys with targeted properties.

\begin{figure}
\begin{center}
\includegraphics[width=0.6\linewidth]{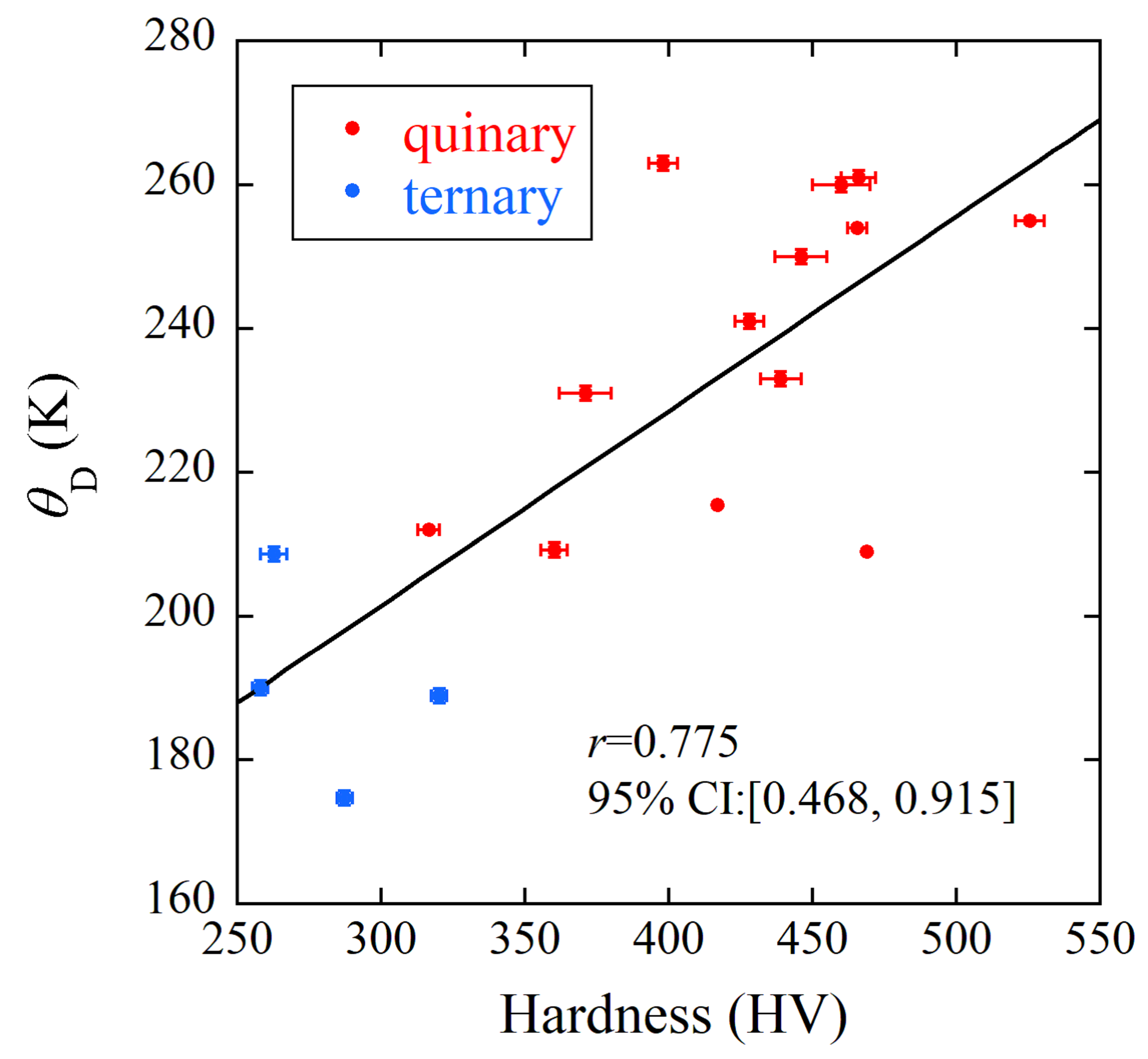}
\caption{\label{fig7} Correlation between $\theta_{D}$ and hardness values.}
\end{center}
\end{figure}

This work has revealed a universal $\lambda_\mathrm{e-p}$-$\theta_\mathrm{D}$ relationship independent of the degree of atomic disorder, indicating that a desired superconducting property can be achieved by controlling $\theta_\mathrm{D}$. 
Here, we propose the simple assessment of $\theta_{D}$ using Vickers microhardness.
Since $\theta_{D}$ represents the maximum phonon frequency and reflects interatomic bonding strength, a higher $\theta_{D}$ implies stronger atomic bonding, which typically results in increased hardness. 
To explore this relationship, we examined the correlation between $\theta_{D}$ and the Vickers microhardness, as shown in Fig.\hspace{1mm}\ref{fig7}, indicating a positive correlation.
Vickers microhardness can be measured rapidly at room temperature and may therefore serve as a practical screening tool for identifying bcc superconducting alloys with desired properties.
Finally, we comment that quinary alloys tend to be harder than ternary alloys. 
The 'four core effects' are well-known distinctive properties of HEAs: high entropy effect, severe lattice distortion, sluggish diffusion, and cocktail effect.
Severe lattice distortion corresponds to solid-solution strengthening caused by pronounced atomic disorder, a feature more characteristic of quinary than ternary alloys. 
Thus, the high-entropy state inevitably induces solid-solution strengthening, leading to increased hardness. 

\section{Summary}
We have verified our previous hypothesis based on the uncertainty principle in bcc HEA superconductors. 
In the equiatomic quinary bcc HEA superconductors, the negative correlation between $\lambda_\mathrm{e-p}$ and $\theta_{D}$ leads to the suppression of $T_\mathrm{c}$ with increasing $\theta_{D}$. 
We previously proposed that this trend can be attributed to the uncertainty principle between energy uncertainty and phonon lifetime; the larger energy uncertainty associated with higher $\theta_{D}$ results in a shortened phonon lifetime, thereby weakening $\lambda_\mathrm{e-p}$ and consequently reducing $T_\mathrm{c}$. 
To examine this hypothesis, the relationship between $\lambda_\mathrm{e-p}$ and $\theta_{D}$ in equiatomic quinary bcc alloys is compared with that in equiatomic ternary bcc alloys. 
It is anticipated that, owing to the large difference in configurational entropy between the two equiatomic systems, the negative correlation between $\lambda_\mathrm{e-p}$ and $\theta_{D}$ would be markedly altered in the ternary alloys. 
However, no significant change is observed when transitioning from the quinary to the ternary system, thus providing limited support for our hypothesis. 
Instead, by incorporating the full set of reported $T_\mathrm{c}$, $\theta_{D}$ and  $\gamma$ data for bcc alloys spanning binary to senary systems, we have identified a universal negative correlation between $\lambda_\mathrm{e-p}$ and $\theta_{D}$ irrespective of the degree of atomic disorder. 
This universal relationship is expected to be valuable for the materials design of bcc alloy superconductors. 
We have also proposed that Vickers microhardness provides an alternative means to assess $\theta_{D}$, and this measurement may serve as a practical tool for the rapid screening of bcc superconducting alloys with desired properties.

\section*{Acknowledgments}
J.K. is grateful for the support provided by the Comprehensive Research Organization of Fukuoka Institute of Technology and a Grant-in-Aid for Scientific Research (KAKENHI) (Grant No. 23K04570). T.N. acknowledges the support from a Grant-in-Aid for Scientific Research (KAKENHI) (Grant No. 24K08236). Y.M. acknowledges the support from a Grant-in-Aid for Scientific Research (KAKENHI) (Grant No. 21H00151), JST-ERATO(Grant No. JPMJER2201), and TMU research fund for young scientist. 

\section*{Data availability statement}   
The data that support the findings of this study are available on reasonable request.

\section*{Author contributions}
Hanabusa Senga: Investigation. Yuto Watanabe: Investigation. Fubuki Iwase: Investigation. Ryo Masuda: Investigation. Daichi Kawahara: Investigation. Toshiki Haruyama: Investigation. Terukazu Nishizaki: Investigation, Formal analysis, Writing - reviewing \& editing. Yoshikazu Mizuguchi: Investigation, Formal analysis, Writing - reviewing \& editing. Jiro Kitagawa: Supervision, Formal analysis, Writing - original draft, Writing - reviewing \& editing. 

\section*{Conflicts of interest}
The authors declare no conflict of interest.

\end{document}